%% file: main.tex
\documentclass[12pt,a4paper]{article}

\usepackage{lineno}  

\usepackage{graphicx}  

\usepackage{xspace}
\usepackage{color}
\usepackage{colortbl}

\usepackage{amsmath}

\usepackage{ifthen} 

\usepackage{graphpap}
\usepackage{rotating}
\usepackage{cancel}

\usepackage{hyperref}
\usepackage[all]{hypcap}

\newboolean{pdflatex}
\setboolean{pdflatex}{false} 
\setboolean{pdflatex}{true} 
\newboolean{uprightparticles}
\setboolean{uprightparticles}{true} 
\usepackage{amssymb}
\usepackage{amsfonts}
\usepackage{upgreek}
\input{lhcb-symbols-def}

\def\dipsi {\ensuremath{{\PJ\mskip -3mu/\mskip -2mu\Ppsi\mskip 2mu}{\PJ\mskip -3mu/\mskip -2mu\Ppsi\mskip 2mu}}\xspace}

\begin{document}


\input{title-LHCb-PAPER}




\pagestyle{plain} 
\setcounter{page}{1}
\pagenumbering{arabic}


%

\input{introduction}

\input{detector}

\input{evsel}

\input{effic}

\input{syst}

\input{xsect}

\input{summary}

\input{acknowledgments}

\input{references}

\end{document}

%% file: lhcb-symbols-def.tex
\def\ux85 {UX85\xspace}

\ifthenelse{\boolean{uprightparticles}}%
{

 \def\Pmu         {\ensuremath{\upmu}\xspace}

 \def\Ppsi        {\ensuremath{\uppsi}\xspace}

 \def\PDelta      {\ensuremath{\Delta}\xspace}                 
 \def\PXi         {\ensuremath{\Xi}\xspace}                 
 \def\PLambda     {\ensuremath{\Lambda}\xspace}                 
 \def\PSigma      {\ensuremath{\Sigma}\xspace}                 
 \def\POmega      {\ensuremath{\Omega}\xspace}                 
 \def\PUpsilon    {\ensuremath{\Upsilon}\xspace}

 \def\PB      {\ensuremath{\mathrm{B}}\xspace}                 
                  
 \def\PD      {\ensuremath{\mathrm{D}}\xspace}

 \def\PJ      {\ensuremath{\mathrm{J}}\xspace}                 
 \def\PK      {\ensuremath{\mathrm{K}}\xspace}

 \def\Pi      {\ensuremath{\mathrm{i}}\xspace}

}
{

 \def\Pmu         {\ensuremath{\mu}\xspace}

 \def\Ppsi        {\ensuremath{\psi}\xspace}                 
                  
 \mathchardef\PDelta="7101
 \mathchardef\PXi="7104
 \mathchardef\PLambda="7103
 \mathchardef\PSigma="7106
 \mathchardef\POmega="710A
 \mathchardef\PUpsilon="7107
                  
 \def\PB      {\ensuremath{B}\xspace}                 
                  
 \def\PD      {\ensuremath{D}\xspace}

 \def\PJ      {\ensuremath{J}\xspace}                 
 \def\PK      {\ensuremath{K}\xspace}

 \def\Pi      {\ensuremath{i}\xspace}

}

\def\mup        {\ensuremath{\Pmu^+}\xspace}
\def\mun        {\ensuremath{\Pmu^-}\xspace} 
\def\mumu       {\ensuremath{\Pmu^+\Pmu^-}\xspace}

\def\kaon  {\ensuremath{\PK}\xspace}
  \def\Kbar  {\kern 0.2em\overline{\kern -0.2em \PK}{}\xspace}

\def\Kz    {\ensuremath{\kaon^0}\xspace}
\def\Kzb   {\ensuremath{\Kbar^0}\xspace}
\def\KzKzb {\ensuremath{\Kz \kern -0.16em \Kzb}\xspace}
\def\Kp    {\ensuremath{\kaon^+}\xspace}
\def\Km    {\ensuremath{\kaon^-}\xspace}

\def\KpKm  {\ensuremath{\Kp \kern -0.16em \Km}\xspace}

  \def\Dbar    {\kern 0.2em\overline{\kern -0.2em \PD}{}\xspace}
\def\D       {\ensuremath{\PD}\xspace}

\def\Dz      {\ensuremath{\D^0}\xspace}
\def\Dzb     {\ensuremath{\Dbar^0}\xspace}
\def\DzDzb   {\ensuremath{\Dz {\kern -0.16em \Dzb}}\xspace}
\def\Dp      {\ensuremath{\D^+}\xspace}
\def\Dm      {\ensuremath{\D^-}\xspace}

\def\DpDm    {\ensuremath{\Dp {\kern -0.16em \Dm}}\xspace}

  \def\Bbar    {\kern 0.18em\overline{\kern -0.18em \PB}{}\xspace}

\def\jpsi     {\ensuremath{{\PJ\mskip -3mu/\mskip -2mu\Ppsi\mskip 2mu}}\xspace}

  \def\Y#1S{\ensuremath{\PUpsilon{(#1S)}}\xspace}

\def\BR         {{\ensuremath{\cal B}\xspace}}

\def\to                 {\ensuremath{\rightarrow}\xspace}

\def\AT#1     {\ensuremath{A_T^{#1}}\xspace}           

\def\C#1      {\ensuremath{\mathcal{C}_{#1}}\xspace}                       
\def\Cp#1     {\ensuremath{\mathcal{C}_{#1}^{'}}\xspace}                    
\def\Ceff#1   {\ensuremath{\mathcal{C}_{#1}^{\mathrm{(eff)}}}\xspace}        
\def\Cpeff#1  {\ensuremath{\mathcal{C}_{#1}^{'\mathrm{(eff)}}}\xspace}       
\def\Ope#1    {\ensuremath{\mathcal{O}_{#1}}\xspace}                       
\def\Opep#1   {\ensuremath{\mathcal{O}_{#1}^{'}}\xspace}                    

\newcommand{\tev}{\ensuremath{\mathrm{\,Te\kern -0.1em V}}\xspace}
\newcommand{\gev}{\ensuremath{\mathrm{\,Ge\kern -0.1em V}}\xspace}
\newcommand{\mev}{\ensuremath{\mathrm{\,Me\kern -0.1em V}}\xspace}
\newcommand{\kev}{\ensuremath{\mathrm{\,ke\kern -0.1em V}}\xspace}
\newcommand{\ev}{\ensuremath{\mathrm{\,e\kern -0.1em V}}\xspace}
\newcommand{\gevc}{\ensuremath{{\mathrm{\,Ge\kern -0.1em V\!/}c}}\xspace}
\newcommand{\mevc}{\ensuremath{{\mathrm{\,Me\kern -0.1em V\!/}c}}\xspace}
\newcommand{\gevcc}{\ensuremath{{\mathrm{\,Ge\kern -0.1em V\!/}c^2}}\xspace}
\newcommand{\gevgevcccc}{\ensuremath{{\mathrm{\,Ge\kern -0.1em V^2\!/}c^4}}\xspace}
\newcommand{\mevcc}{\ensuremath{{\mathrm{\,Me\kern -0.1em V\!/}c^2}}\xspace}

\def\gsim{{~\raise.15em\hbox{$>$}\kern-.85em
          \lower.35em\hbox{$\sim$}~}\xspace}
\def\lsim{{~\raise.15em\hbox{$<$}\kern-.85em
          \lower.35em\hbox{$\sim$}~}\xspace}

\def\tell1  {TELL1\xspace}
\def\ukl1   {UKL1\xspace}



%% file: title-LHCb-PAPER.tex

\begin{titlepage}
\pagenumbering{roman}

\vspace*{-1.5cm}
\centerline{\large EUROPEAN ORGANIZATION FOR NUCLEAR RESEARCH (CERN)}
\vspace*{1.5cm}
\hspace*{-0.5cm}
\begin{tabular*}{\linewidth}{lc@{\extracolsep{\fill}}r}
\ifthenelse{\boolean{pdflatex}}
{\vspace*{-2.7cm}\mbox{\!\!\!\includegraphics[width=.14\textwidth]{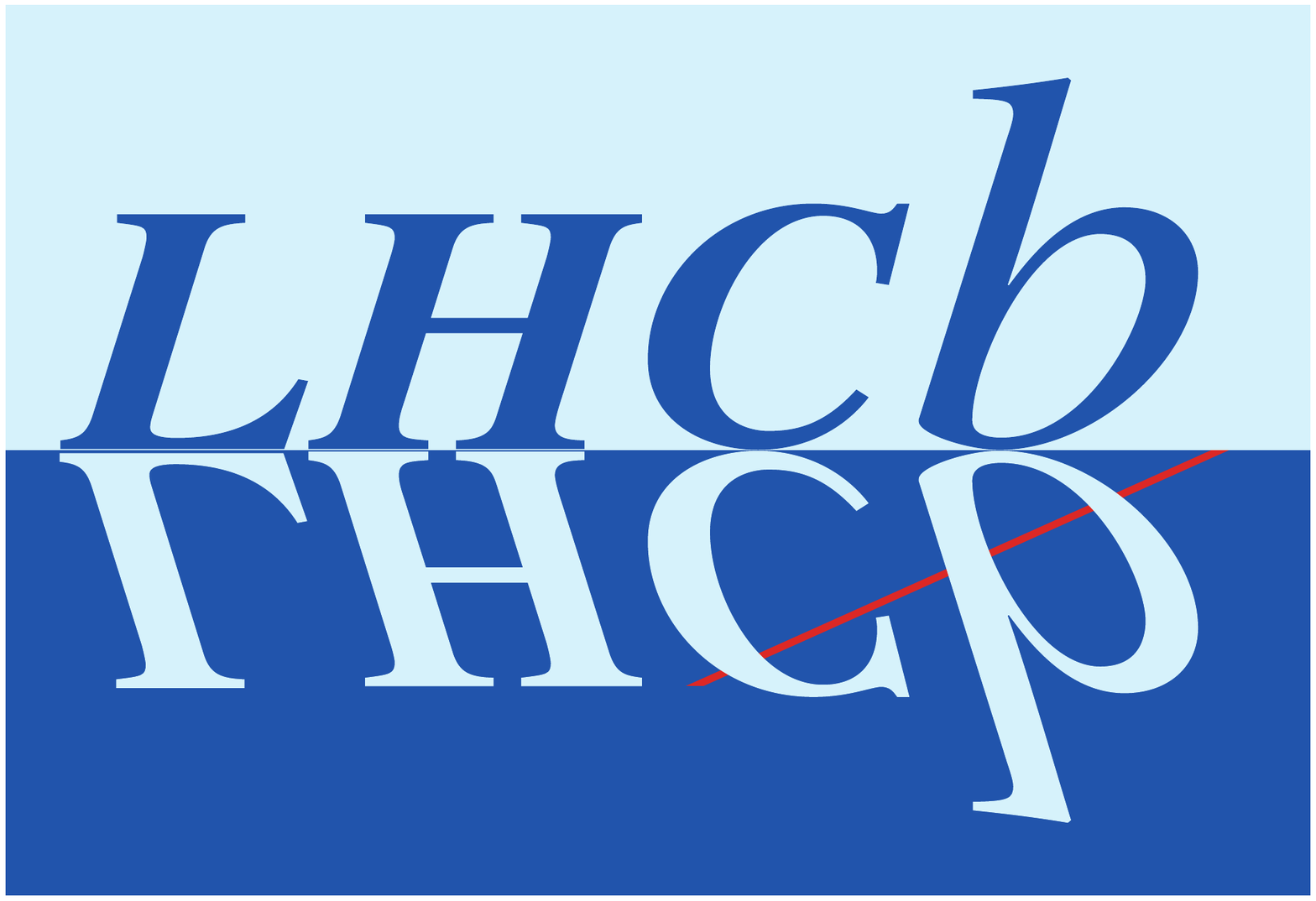}} & &}%
{\vspace*{-1.2cm}\mbox{\!\!\!\includegraphics[width=.12\textwidth]{lhcb-logo.eps}} & &}%
\\
 & & LHCb-PAPER-2011-013 \\  
 & & CERN-PH-EP-2011-135 \\  
 & & \today \\ 
 & &        \\ 
\end{tabular*}

\vspace*{4.0cm}

{\bf\boldmath\huge
\begin{center}
  Observation of \jpsi~pair production 
  in pp~collisions at $\sqrt{s}=7~\mathrm{TeV}$
\end{center}
}

\vspace*{2.0cm}

\begin{center}
The LHCb Collaboration
\footnote{Authors are listed on the following pages.}
\end{center}

\vspace{\fill}

\begin{abstract}
  \noindent
  The production of \jpsi~pairs in proton-proton collisions 
  at a centre-of-mass energy of $7~\mathrm{TeV}$  
  has been observed using an integrated luminosity of 
  $37.5~\mathrm{pb}^{-1}$ collected  with the LHCb detector.
  The production cross-section for pairs
  with both \jpsi~in the rapidity range 
  \mbox{ $2<y^{\jpsi}<4.5$}
  and transverse momentum
  \mbox{$p_{\mathrm{T}}^{\jpsi}<10~\mathrm{GeV}/c$} 
  is
  \begin{equation*}
    \sigma^{\dipsi} 
    = 
    5.1\pm1.0\pm1.1~\mathrm{nb},
  \end{equation*}
  where the first uncertainty is statistical 
  and the second systematic.
\end{abstract}

\vspace*{2.0cm}
\vspace{\fill}

\begin{center}
(Submitted to Physics Letters B.)
\end{center}

\end{titlepage}


\newpage
\setcounter{page}{2}
\mbox{~}
\newpage

\input{LHCb_authorlist.tex}

\cleardoublepage

%% file: LHCb_authorlist.tex
\noindent
R.~Aaij$^{23}$, 
B.~Adeva$^{36}$, 
M.~Adinolfi$^{42}$, 
C.~Adrover$^{6}$, 
A.~Affolder$^{48}$, 
Z.~Ajaltouni$^{5}$, 
J.~Albrecht$^{37}$, 
F.~Alessio$^{37}$, 
M.~Alexander$^{47}$, 
G.~Alkhazov$^{29}$
% ...
\\

\noindent 
{\it \footnotesize 
% ...
$ ^{5}$Clermont Universit\'{e}, Universit\'{e} Blaise Pascal, CNRS/IN2P3, LPC, Clermont-Ferrand, France\\
$ ^{6}$CPPM, Aix-Marseille Universit\'{e}, CNRS/IN2P3, Marseille, France\\
% ...
$ ^{23}$Nikhef National Institute for Subatomic Physics, Amsterdam, Netherlands\\
% ...
$ ^{29}$Petersburg Nuclear Physics Institute (PNPI), Gatchina, Russia\\
% ...
$ ^{36}$Universidad de Santiago de Compostela, Santiago de Compostela, Spain\\
$ ^{37}$European Organization for Nuclear Research (CERN), Geneva, Switzerland\\
% ...
$ ^{42}$H.H. Wills Physics Laboratory, University of Bristol, Bristol, United Kingdom\\
% ...
$ ^{47}$School of Physics and Astronomy, University of Glasgow, Glasgow, United Kingdom\\
$ ^{48}$Oliver Lodge Laboratory, University of Liverpool, Liverpool, United Kingdom\\
% ...
}

%% file: introduction.tex
%

\section{Introduction}
\label{sec:Introduction}
The mechanism of heavy quarkonium production 
is a long-standing problem in QCD. 
An~effective field theory, non-relativistic QCD (NRQCD),
provides the foundation for much of the current theoretical work. 
According to NRQCD, the production of heavy quarkonium 
factorizes into two steps:  a heavy quark-antiquark pair is first created 
perturbatively at short distances and subsequently evolves
non-perturbatively into quarkonium at long distances. 
The NRQCD calculations depend
on the colour-singlet (CS) and colour-octet (CO) matrix elements, 
which account for the probability of a heavy quark-antiquark pair 
in a particular colour state to evolve into heavy quarkonium. 

Leading order (LO) calculations in the 
CS model~\cite{Kartvelishvili:1978id,Berger:1980ni,Baier:1981uk} were 
first used to describe experimental data. 
However, they underestimate the~observed
cross-section for single \jpsi~production at high $p_{\mathrm{T}}$ at the 
Tevatron~\cite{Abe:1992ww}. To~resolve this
discrepancy the CO mechanism was introduced~\cite{Braaten:1994vv}. 
The corresponding matrix elements were determined from the
large-$p_{\mathrm{T}}$ data as the CO cross-section falls more slowly
than the CS one. However, recent 
calculations~\cite{Campbell:2007ws,Gong:2008sn,artoisenet:2008,lansberg:2009} 
close the gap between the CS predictions and the experimental
data~\cite{Brambilla} reducing the need for large CO contributions.
Thus, further experimental tests are needed. Pair production of quarkonium 
can cast light on this problem as this process depends heavily 
on the production mechanism.  
For~both the CS and CO models, 
contributions  from double parton 
scattering~\cite{Kom:2011bd,Baranov:2011ch,Novoselov:2011ff}
could potentially be significant.

The only observation of charmonia pair production in 
hadronic collisions to date was by the NA3 collaboration, 
who found evidence for \jpsi~pair production in multi-muon events
in pion-platinum interactions at pion momenta of~150~and 
280~GeV$/c$~\cite{Badier:1982ae}
and in proton-platinum interactions at a proton momentum 
of~400~GeV$/c$~\cite{Badier:1985ri}.
The cross-section ratio $\sigma^{\dipsi}/\sigma^{\jpsi}$
was measured to be  $(3\pm1)\times10^{-4}$ for
pion-induced production, where $\sigma^{\jpsi}$ is the
inclusive \jpsi~production cross-section. At~NA3 energies the main
contribution to the \jpsi~pair cross-section arises from 
the~quark-antiquark annihilation
channel~\cite{Kartvelishvili:1984ur}. This is not the case for proton-proton 
collisions at the~LHC, where the gluon-gluon fusion process 
dominates~\cite{Humpert:1983yj,Kiselev:1988mc}.

Theoretical calculations based on the LO production of CS-states
predict that the total cross-section for $\jpsi$-pair production 
in proton-proton interactions at $\sqrt{s}=7$~TeV is equal 
to $24$~nb~\cite{Qiao:2009kg,Berezhnoy:2011xy}. 
These calculations take into account 
\dipsi, $\jpsi\Ppsi(2S)$ and 
$\Ppsi(2S)\Ppsi(2S)$~production 
but do not include the possible 
contribution from double parton scattering. 
In the rapidity interval $2.0<y^{\jpsi}<4.5$, 
relevant to the LHCb experiment,
the expected value is $4~\mathrm{nb}$ with an uncertainty of around~30\%. 
At small invariant masses of the \jpsi~pair a~tetraquark state, 
built from four $c$-quarks, may exist~\cite{Berezhnoy:2011xy} and 
would be visible as a narrow resonance in the mass spectrum.

%% file: detector.tex
%

\section{The LHCb detector and dataset }
\label{sec:detector}
The LHCb detector is a forward spectrometer \cite{LHCb} providing 
charged particle reconstruction in the pseudorapidity range 
$1.9<\eta<4.9$.  The detector elements are 
placed along the beam line of the LHC starting with the Vertex Locator, a
silicon strip device that surrounds the proton-proton interaction region. 
This reconstructs precisely the locations of interaction vertices, 
the locations of decays of long-lived hadrons and contributes 
to the measurement of track momenta. Other detectors 
used to measure track momenta
comprise a large area silicon strip detector located 
upstream of a dipole magnet with bending power around 4~Tm
and a combination of silicon strip detectors and straw 
drift-tubes placed downstream.
Two ring imaging Cherenkov detectors are used to identify charged hadrons. 
Further downstream, an electromagnetic calorimeter is used for photon and
electron identification, followed by a hadron calorimeter and a muon 
system consisting of alternating layers of iron and chambers 
(MWPC and triple-GEM) that distinguishes muons from hadrons. 
The calorimeters and muon system provide the capability of
first-level hardware triggering.

The LHCb trigger system consists of three levels. The first level
(L0) is designed to reduce the LHC bunch crossing rate of 40 MHz to a 
maximum of 1 MHz, at which the complete detector is read out. 
This is the input to the first stage of the software trigger, 
which performs a partial event reconstruction to confirm or discard 
the L0 trigger decision. The second stage of the software trigger 
performs a full event reconstruction to further discriminate 
signal events from other $\mathrm{pp}$~collisions. To avoid that a few events 
with high occupancy dominate the CPU time, a set of global event cuts 
is applied on the hit multiplicities of each sub-detector used 
by the pattern recognition algorithms. 
These cuts were chosen to reject high-multiplicity events with a large number 
of $\mathrm{pp}$~interactions with minimal loss of luminosity.

The data used for this analysis comprise an integrated luminosity 
of~$37.5~\mathrm{pb}^{-1}$ of~$\mathrm{pp}$~collisions at a 
centre-of-mass energy of~7~\tev  collected by the LHCb experiment 
between July and November 2010. 
This number includes the dead-time of trigger and 
data acquisition systems.
During this period all detector components were fully
operational and in a stable condition.
The mean number of visible proton-proton collisions
per bunch crossing was up~to~2.5.

The simulation samples used are based on 
the {\sc{Pythia}}~6.4 generator~\cite{pythia} configured 
with the parameters detailed in Ref.~\cite{belyaev}. 
The {\sc{EvtGen}}~\cite{evtgen} and  {\sc{Geant4}}~\cite{geant} packages are used 
to generate hadron decays and  simulate interactions in the detector, 
respectively. Prompt charmonium production is generated in 
{\sc{Pythia}} according to the  leading order CS and CO mechanisms.

%% file: evsel.tex
%

\section{Event selection and signal yield}
\label{sec:EventSelection}
In this analysis the \jpsi is reconstructed through its decay into a pair of muons. 
Events with at least four muons are selected.
$\jpsi\to\mumu$~candidates are formed from pairs of oppositely-charged 
particles identified as muons that each have a transverse momentum greater 
than $650~\mathrm{MeV}/c$ and that originate from a common vertex. Track
quality is ensured by requiring that the $\chi^2_{\rm{tr}}/\mathrm{ndf}$ provided by the
track fit is less than five. Well identified muons are selected by requiring that
the difference in logarithms of the global likelihood of the muon hypothesis, 
provided by the particle identification detectors~\cite{Powell:2010zz}, 
with respect to the hadron hypothesis, 
$\Delta \ln \mathcal{L}^{\mu-\mathrm{h}}$, be greater than zero. 
To suppress the contribution from duplicate particles created by the 
reconstruction procedure, if two muon candidates have a symmetrized 
Kullback-Leibler divergence~\cite{Kullback}
less than 5000, only the particle with
the best track fit is considered.

Selected \mumu~candidates with an invariant mass in the range
\mbox{$3.0<m_{\mumu}<3.2~\mathrm{GeV}/c^2$} are paired to form 
$(\mumu)_{1}(\mumu)_{2}$ combinations. A~fit
of the four-muon candidate is performed~\cite{Hulsbergen}
that requires the four 
tracks to be consistent with originating from a common vertex and that
this vertex is compatible with one of the reconstructed $\mathrm{pp}$~collision
vertices.  To reject background where two \jpsi~candidates originate 
from different $\mathrm{pp}$~collisions,
the reduced $\chi^2$ of this fit, 
$\chi^2_{\mathrm{}}/\mathrm{ndf}$, 
is required to be less than five.

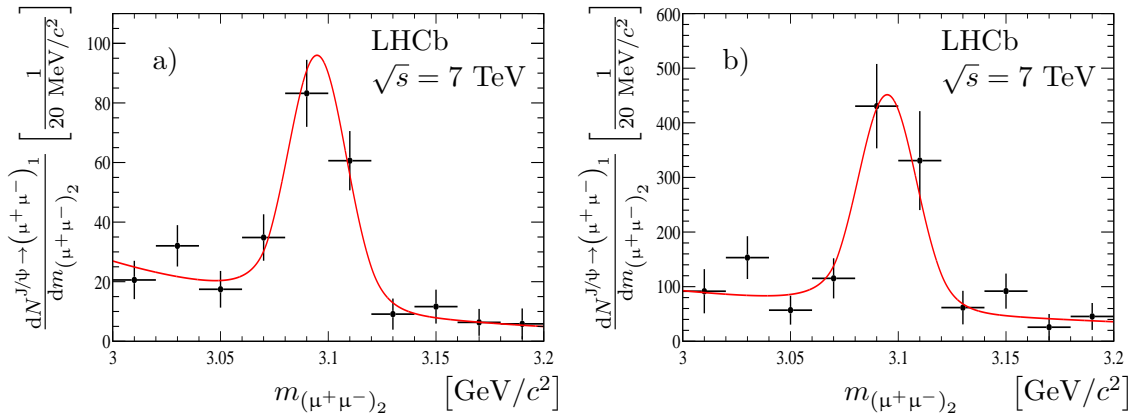
\begin{figure}[htb]
  \setlength{\unitlength}{1mm}
  \centering
  \begin{picture}(150,55)
    %
    \put(5,0){
      \includegraphics*[width=70mm,height=55mm,%
      ]{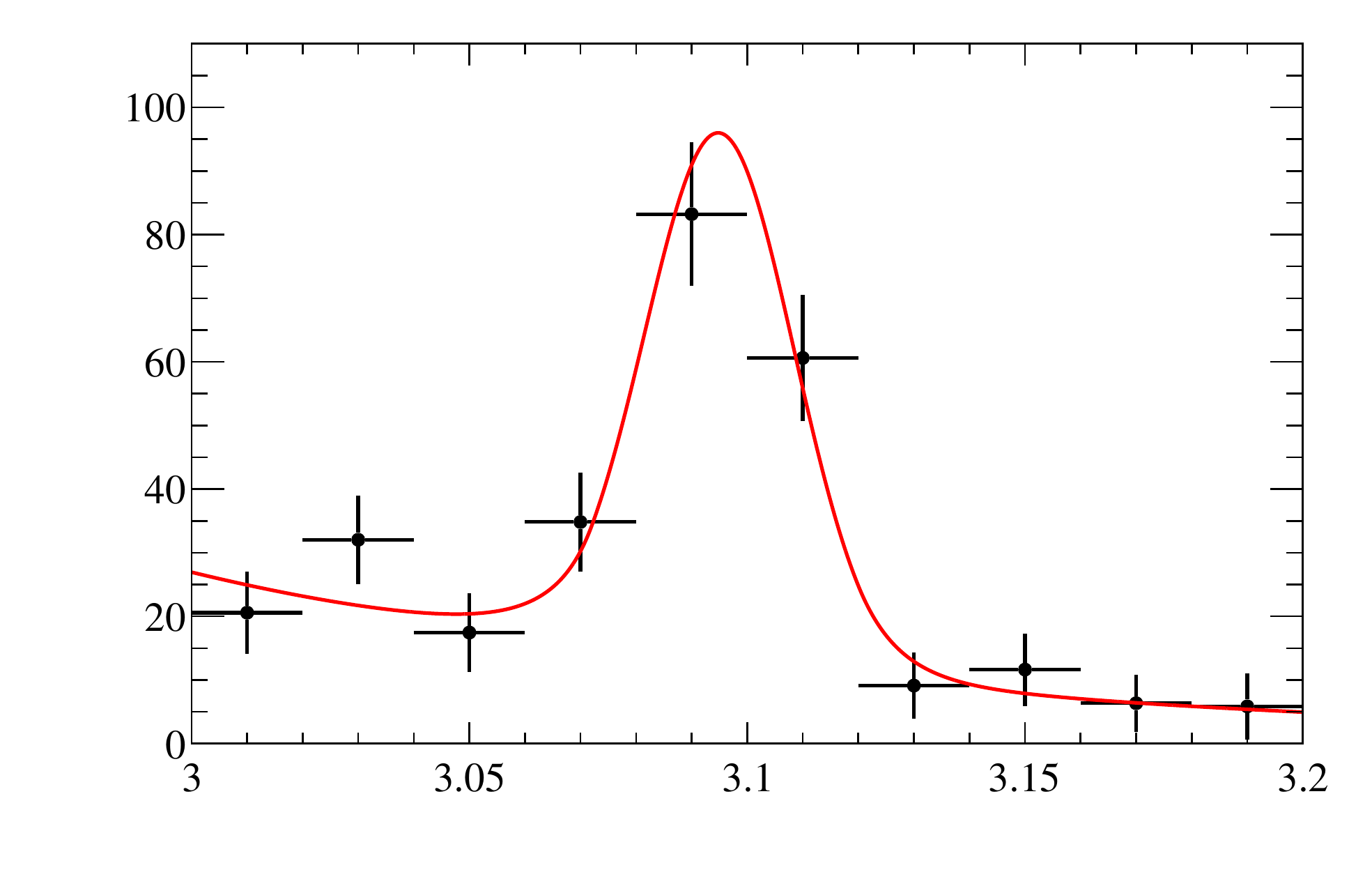}
    }
    \put(80,0){
      \includegraphics*[width=70mm,height=55mm,%
      ]{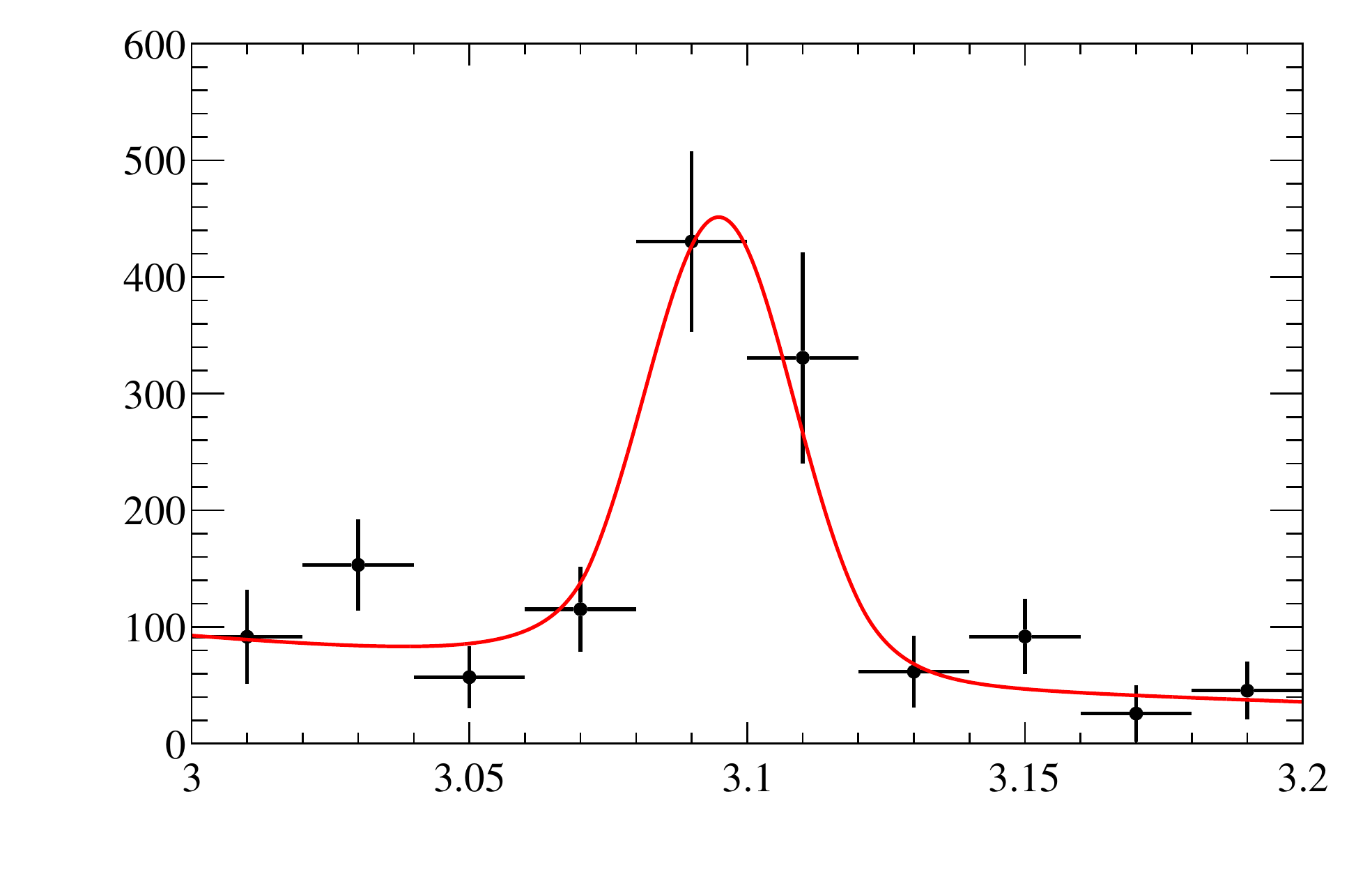}
    }
    \put(36,  1){ \small $m_{\left(\mumu\right)_2}$ }
    \put(111, 1){ \small $m_{\left(\mumu\right)_2}$ }
    \put(58,  1){ \small $\left[ \mathrm{GeV}/c^2 \right]$}
    \put(133, 1){ \small $\left[ \mathrm{GeV}/c^2 \right]$}
    \put(47,45) {  %
      \small
      $\begin{array}{l}                            %
	\text{LHCb}                   \\           %
        \sqrt{s} = 7~\mathrm{TeV}         %
      \end{array}$ } 
    \put(122,45) {  %
      \small
      $\begin{array}{l}                            %
	\text{LHCb}                   \\           %
        \sqrt{s} = 7~\mathrm{TeV}         %
      \end{array}$ } 
    \put(1,10) { 
      \small 
      \begin{sideways}%
	$\frac{ \mathrm{d}N^{\jpsi\to\left(\mumu\right)_1}}{\mathrm{d}m_{\left(\mumu\right)_2}} \left[ \frac{1}{20~\mathrm{MeV}/c^2}\right]$
      \end{sideways} 
    }
    \put(76,10) { 
      \small 
      \begin{sideways}%
	$\frac{ \mathrm{d}N^{\jpsi\to\left(\mumu\right)_1}}{\mathrm{d}m_{\left(\mumu\right)_2}} \left[ \frac{1}{20~\mathrm{MeV}/c^2}\right]$
      \end{sideways} 
    }
    \put(20,45) {  \small a) }
    \put(95,45) {  \small b) } 
  \end{picture}
  \caption {
    The fitted yields of $\jpsi \to \left(\mumu\right)_1$ 
    in bins of $(\mumu)_2$  invariant mass: 
   (a)~the raw signal yield observed in the data;
   (b)~the efficiency corrected yield (Sect.~\ref{sec:CrossSection}). 
    The~result of a fit with a double-sided Crystal Ball function 
    for the signal and an exponential background is superimposed.
  }
  \label{fig:dipsi_raw}
\end{figure}

The number of events with two \jpsi~mesons is extracted from 
the single \jpsi~mass spectra.
The invariant mass distributions of the first muon pair 
are obtained in bins of the invariant mass of 
the second pair.\footnote{ The \mumu~pair with lower transverse 
momentum is chosen  to be the first pair.} 
The single \jpsi~mass spectrum is modelled empirically using simulated events. 
This exhibits non-Gaussian tails on either side of the peak.
The tail on the left-hand side is dominated by 
radiative effects in \jpsi~decay, while the right-hand side tail is due 
to non-Gaussian effects in the reconstruction. 
The~shape of the distribution is described by a function 
that is similar to a Crystal~Ball function~\cite{Gaiser,Skwarnicki}, 
but with the power-law tails on both sides 
of the core Gaussian component. 
The position of the \jpsi~peak, the  effective mass resolution 
and the tail parameters of this double-sided Crystal Ball function
are fixed to the values determined from an analysis of the signal 
shape in the inclusive \jpsi~sample. Combinatorial background is
modelled using an exponential function.
This model is used to extract the yield of 
$\jpsi \to \left(\mup \mun\right)_1$  in bins 
of the $(\mumu)_{2}$  invariant  mass. 
The extracted yield is shown  in 
Fig.~\ref{fig:dipsi_raw}a together with the result of a 
fit according to the model described above. 
The yield of events with double \jpsi production given 
by the fit is 
$N^{\dipsi} = 141\pm19$, 
where the 
statistical significance of this signal exceeds $6\sigma$.
A~fit with position and resolution of the signal peak left free was 
also performed and gave consistent results.

Studies of single \jpsi~production indicate that the detector
acceptance and efficiency is high for the fiducial range
$2<y^{\jpsi}<4.5$ and $p_{\mathrm{T}}^{\jpsi}<10~\mathrm{GeV}/c$. The
raw yield of events with both \jpsi~mesons within this range is
$139\pm18$.  
The yield of events with both \jpsi~mesons in the fiducial 
range and explicitly triggered  
by one of the \jpsi~candidates
through the single muon or 
dimuon trigger lines~\cite{Aaij:2011rj},
is found to be  $116\pm16$. 
This sample is considered for the 
determination of the production cross-section.

The contribution to the yield from  the pileup of two interactions each 
producing a~single \jpsi~meson is estimated using simulation together with the
measured \jpsi~production cross-section~\cite{Jpsi}. This study shows that 
for the 2010 data-taking conditions the~background  from this source is expected to 
be less than 1.5 events and hence can be neglected. In a similar way the contribution 
to the yield from events with \jpsi~mesons originating from the decays of 
beauty hadrons is found to be negligible.

%% file: effic.tex
%

\section{Efficiency evaluation}
\label{sec:Efficiency}

The per-event efficiency for a \jpsi-pair event,
$\varepsilon^{\mathrm{tot}}_{\dipsi}$, 
is decomposed into three factors,
\begin{equation}
  \varepsilon^{\mathrm{tot}}_{\dipsi}
  =
  \varepsilon^{\mathrm{reco}}_{\dipsi} \times 
  \varepsilon^{\mathrm{\mu ID}}_{\dipsi} \times 
  \varepsilon^{\mathrm{trg}}_{\dipsi}, 
\label{eq:total_eff}
\end{equation}
where $\varepsilon^{\mathrm{reco}}_{\dipsi}$ is the product 
of the (geometrical) acceptance with 
reconstruction and selection efficiency, 
$\varepsilon^{\mathrm{\mu ID}}_{\dipsi}$ is the~efficiency 
for muon identification and  $\varepsilon^{\mathrm{trg}}_{\dipsi}$ is 
the trigger efficiency for reconstructed and selected events. 

The efficiency for the acceptance, reconstruction and selection for 
the two \jpsi~mesons is factorized into the product of  efficiencies 
for the first and second \jpsi,
\begin{equation}
\varepsilon^{\mathrm{reco}}_{\dipsi}= 
\varepsilon^{\mathrm{reco}}_{\jpsi} \left( p^{\jpsi_1}_{\mathrm{T}}, y^{\jpsi_1} , | \cos \vartheta^{*}_{\jpsi_1} | \right) 
\times 
\varepsilon^{\mathrm{reco}}_{\jpsi} \left( p^{\jpsi_2}_{\mathrm{T}}, y^{\jpsi_2} , | \cos \vartheta^{*}_{\jpsi_2} | \right).
\label{eq:eff_factorization}
\end{equation}
The single \jpsi~efficiency $\varepsilon^{\mathrm{reco}}_{\jpsi}$ is 
a function of the~transverse momentum $p_{\mathrm{T}}$,  
rapidity $y$  and $\left| \cos \vartheta^*\right|$,  
where $\vartheta^{*}$ is the angle 
between the \mup~momentum  in the \jpsi~centre-of-mass frame and 
the \jpsi~flight direction in the~laboratory frame. 
It is evaluated using simulation.
The~validity of the factorization hypothesis of Eq.~\eqref{eq:eff_factorization}
is checked with simulation and based on these studies a correction factor of 
0.975~is applied to~$\varepsilon^{\mathrm{reco}}_{\dipsi}$. 
For the simulated data of single prompt \jpsi production the cut on
the muon likelihood  is not applied and that on 
$\chi^2_{\mathrm{}}(\dipsi)/\mathrm{ndf}$ 
is replaced with a~similar cut on the single \jpsi, 
\mbox{$\chi^2_{\mathrm{}}(\jpsi)/\mathrm{ndf}<5$}.
The efficiency of the cut on 
$\chi^2_{\mathrm{}}/\mathrm{ndf}$ 
is estimated from
the data and compared to the~simulation.  Based on these studies a correction factor 
of 1.026 is applied to $\varepsilon^{\mathrm{reco}}_{\dipsi}$, 
and a~systematic uncertainty of 3\%~is assigned.
The efficiency $\varepsilon^{\mathrm{reco}}_{\jpsi}$ is also corrected by 
a~factor $1.024\pm0.011$~\cite{Jpsi}, that accounts for the ratio of the
reconstruction efficiency of the~muon detector observed
in the data compared to the simulation. 

The muon identification efficiency is extracted from the analysis 
of the inclusive \jpsi~sample. 
Two efficiencies are evaluated:
the single muon identification 
efficiency $\varepsilon_{\mu}^{\mathrm{\mu ID}}$~and 
the \jpsi~efficiency $\varepsilon_{\jpsi}^{\mathrm{\mu ID}}$. 
Both are measured as a function of the value of the cut made on 
$\Delta \ln \mathcal{L}^{\mu-\mathrm{h}}$. 
The squared efficiency 
$(\varepsilon_{1\mu}^{\mu\mathrm{ID}})^2$ and 
$\varepsilon_{\jpsi}^{\mu\mathrm{ID}}$ are found to 
be equal to better than one per mille. 
The value of 
$\varepsilon^{\mathrm{\mu ID}}_{\dipsi}= (\varepsilon^{\mathrm{\mu ID}}_{\jpsi})^2=(91.0 \pm 0.1)\%$ has 
been used as a global factor for the evaluation of the total
efficiency using Eq.\ \eqref{eq:total_eff}.  As a cross-check, 
the efficiency of the muon identification for 
\jpsi~pairs has been estimated from
the signal itself. Though statistically limited, the value
is consistent with that given above.

The trigger efficiency is calculated  to be 
\begin{equation*}
\varepsilon^{\mathrm{trg}}_{\dipsi}= 1 -     
\left(1-\varepsilon^{\mathrm{trg}}_{\jpsi} \left( p^{\jpsi_1}_{\mathrm{T}}, y^{\jpsi_1} \right) \right) 
\times  
\left(1-\varepsilon^{\mathrm{trg}}_{\jpsi} \left( p^{\jpsi_2}_{\mathrm{T}}, y^{\jpsi_2} \right) \right).
\end{equation*}
The trigger efficiency for a single \jpsi,
$\varepsilon^{\mathrm{trg}}_{\jpsi}$, 
is determined directly on data from 
the inclusive prompt \jpsi sample as 
a function of $p_{\mathrm{T}}$  and rapidity $y$. 
The efficiency is determined by classifying an event 
which would also have been triggered without 
the \jpsi~as TIS (Trigger Independent of Signal), 
and/or classifying the event where the \jpsi~alone is sufficient 
to trigger the event as a TOS(Trigger On Signal) event~\cite{TisTos,K0S}. 
The LHCb trigger system records all the information needed 
for such classification. Events can be classified as TIS and TOS 
simultaneously (TIS\&TOS),  which allows the extraction of the trigger 
efficiency relative to the off-line  reconstructed and selected 
events from data alone
\begin{equation}
 \varepsilon^{\mathrm{trg}}_{\jpsi} 
= 
\frac
{N^{\mathrm{TIS\&TOS}}}
{N^{\mathrm{TIS}}}, 
\nonumber
\end{equation}
where 
$N^{\mathrm{TIS}}$ is the number of TIS events, and $N^{\mathrm{TIS\&TOS}}$ is 
the number of events that are simultaneously TIS and TOS.
The method has been cross checked using Monte Carlo simulation.

The effect of the global event cuts applied in the trigger has 
been studied in detail for~inclusive \jpsi~events in Ref.~\cite{Jpsi}. Since 
the~subdetector hit multiplicity observed in single 
and double \jpsi events is similar, the~efficiency of the global event cuts, $(93\pm2)\%$,  
is taken from that study and applied to~$\varepsilon^{\mathrm{trg}}_{\dipsi}$.   

For selected \jpsi-pair events the mean value of  
$\varepsilon^{\mathrm{reco}}_{\dipsi}$ is 31\% and it varies from 
0.8\% to 70\%. The mean value for $\varepsilon^{\mathrm{trg}}_{\dipsi}$  
is 85\% and it varies from 61\% to 93\%.

%% file: syst.tex
%

\section{Systematic uncertainties}
\label{sec:Systematics}

Systematic uncertainties affecting the cross-section measurement have
been evaluated properly taking correlations into account where appropriate. 
The dominant source of systematic uncertainty is due to the knowledge
of the track-finding efficiency. An uncertainty of  4\% per track is
assigned based on studies comparing the reconstruction 
efficiency in data and simulation using a tag and
probe approach~\cite{K0S}. 

The second major source of uncertainty is due to the evaluation of the
trigger efficiency. The method discussed in Sect.~\ref{sec:Efficiency} has 
been cross-checked in several ways, in particular, by using events 
triggered by the first or second \jpsi~only. The observed differences lead 
to the assignment of an 8\% systematic uncertainty.

A further source of uncertainty is the~determination of the per-event efficiency 
defined by Eq.~\eqref{eq:total_eff}. This is estimated to be $3\%$ by
varying the uncertainties of the various factors entering into  Eq.~\eqref{eq:total_eff}.

The unknown polarization of \jpsi~mesons affects the acceptance, 
reconstruction and selection efficiency
$\varepsilon^{\mathrm{reco}}_{\jpsi}$ \cite{Jpsi}. 
In this analysis the effect is reduced by explicitly taking into
account the dependence of the acceptance on $\varepsilon^{\mathrm{reco}}_{\jpsi}$ on $\left|
  \cos\vartheta^*\right|$ in the efficiency determination (Eq.~\ref{eq:eff_factorization}). 
The remaining dependence
results in a systematic uncertainty of 5\%~per~\jpsi.

Additional systematic uncertainties arise due to 
the~difference observed between the data and simulation for the behaviour of the 
cut on 
$\chi^2_{\mathrm{}}$~($3\%$), 
the global event cuts ($2\%$), 
and uncertainty of 1.1\% per \jpsi associated 
with  the efficiency for muon identification,
as discussed in Sect.~\ref{sec:Efficiency}.  
The systematic uncertainties associated with 
the other selection criteria and the \jpsi~lineshape 
parametrization are  negligible.

\begin{table}[thb]
  \centering
  \caption{
   Relative systematic uncertainties on the cross-section measurement. 
   The total uncertainty is calculated as the quadratic sum of the individual 
   components.
  } \label{tab:syst}
  \vspace*{3mm}
  \begin{tabular*}{0.9\textwidth}{@{\hspace{10mm}}l@{\extracolsep{\fill}}c@{\hspace{10mm}}}
    \hline       
    Source &  Systematic uncertainty $\left[\%\right]$\\
    \hline    
    Track finding efficiency  
    & $4\times4$      \\
    Trigger efficiency
    & 8              \\ 
    Per-event efficiency
    & 3              \\ 
    \jpsi~polarization 
    & $2\times5$      \\ 
    Data/simulation difference for $\chi^2_{\mathrm{}}/{\mathrm{ndf}}$ 
    & 3               \\
    Global event cuts 
    & 2               \\
    Muon identification  
    & $2\times1.1$              \\
    Luminosity 
    & 3.5             \\
    $\jpsi\to\mumu$ branching ratio 
    & $2\times1$    \\
    \hline
    Total 
    & 21              \\
    \hline    
  \end{tabular*}
\end{table}

The luminosity was measured at specific periods during the data taking using both
van~der~Meer scans~\cite{VanDerMeer} and a beam-gas imaging method~\cite{BeamProfile}. 
The instantaneous luminosity determination is then based on a continuous recording 
of the multiplicity of tracks reconstructed in Vertex Locator, which has been 
normalized to the absolute luminosity scale. Consistent results are found for 
the absolute luminosity scale with a precision of 3.5\%, dominated by the beam current 
uncertainty~\cite{Lumi,Lumi2}.

The relative systematic uncertainties are summarized in Table~\ref{tab:syst}, where  
the total uncertainty is defined as the quadratic sum of the individual components.

%% file: xsect.tex
%

\section{Cross-section determination }
\label{sec:CrossSection}

The efficiency-corrected yield for events with 
both \jpsi~candidates  in the fiducial region is extracted
using the  procedure discussed in
Sect.~\ref{sec:EventSelection}. 
To account for the efficiency 
a~weight $\omega$, defined as
\begin{equation*}
  \omega^{-1} = 
  \varepsilon^{\mathrm{tot}}_{\dipsi}
\end{equation*}
where $\varepsilon^{\mathrm{tot}}_{\dipsi}$ is the total efficiency defined 
in Eq.~\eqref{eq:total_eff}, is applied to each candidate in the~sample.

The corrected yield of $\jpsi \to \left(\mumu\right)_1$ in bins of
$(\mumu)_2$  invariant mass is shown in Fig.~\ref{fig:dipsi_raw}b. 
As previously described, to extract the yield a fit with a double-sided 
Crystal Ball function for the~signal, together with an 
exponential function for the~background component, is performed. Again,
the position of the \jpsi\ peak and the 
effective mass resolution are fixed to the values found in the~inclusive 
\jpsi\ sample. The event yield after the efficiency correction is
\begin{equation*}
    N^{\mathrm{corr}}_{\dipsi}   = 672 \pm 129.
\end{equation*}

The cross-section for double \jpsi\ production in the fiducial range $2<y^{\jpsi}<4.5$ and 
$p_{\mathrm{T}}^{\jpsi}<10~\mathrm{GeV}/c$ is computed as
\begin{equation*}
  \sigma^{\dipsi} = 
  \frac{ N^{\mathrm{corr}}_{\dipsi}}
       {\mathcal{L}\times
	 \BR_{\mumu}^2}, 
\end{equation*}
where 
$\mathcal{L}=37.5\pm1.3~\mathrm{pb}^{-1}$ is the integrated 
luminosity
and 
$\BR_{\mumu}=(5.93\pm0.06)\%$~\cite{PDG}
is the $\jpsi\to\mumu$ 
branching ratio.
The~result is
\begin{equation*}
  \sigma^{\dipsi} = 
  5.1       
  \pm 1.0   
  \pm 1.1~\mathrm{nb},  
  \label{eq:xsect}
\end{equation*}
where the first uncertainty is statistical and the second systematic.

Using the measured prompt \jpsi production 
cross-section in the same fiducial region~\cite{Jpsi} and 
taking  into account the correlated uncertainties, 
the ratio of cross-sections $\sigma^{\dipsi}/\sigma^{\jpsi}$ 
is calculated to be 
\begin{equation*}
  \sigma^{\dipsi}/\sigma^{\jpsi} = 
\left ( 
        5.1 
    \pm 1.0 
    \pm 0.6 
    \,^{+1.2}_{-1.0} \right) \times 10^{-4},
\end{equation*}
where the first error is statistical, the second systematic 
and  the third is due to the unknown 
polarization of the prompt \jpsi and \jpsi from pair 
production.

\begin{figure}[htb]
  \setlength{\unitlength}{1mm}
  \centering
  \begin{picture}(140,90)
    \put(0,-1){
      \includegraphics*[width=140mm,
      ]{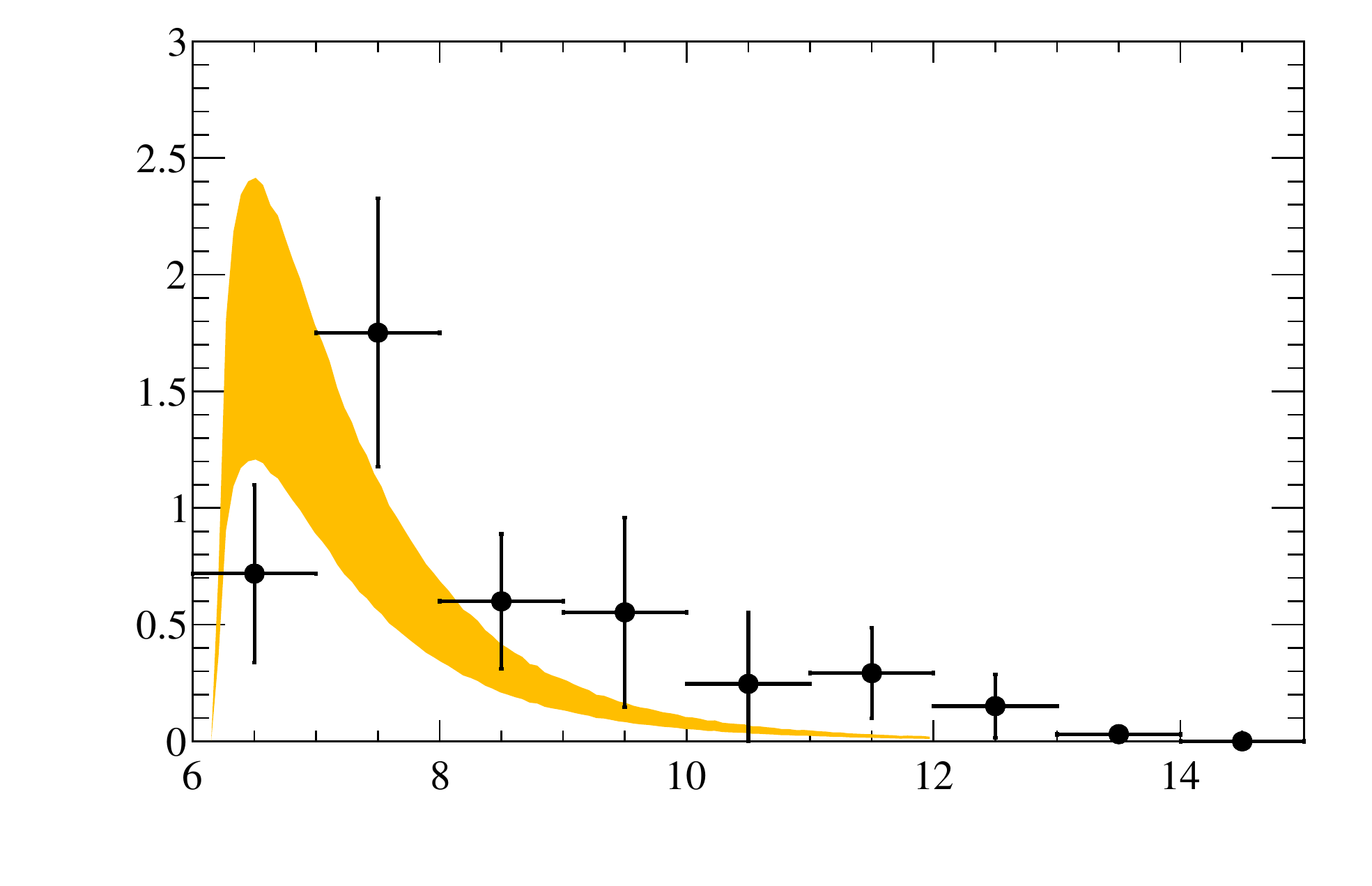}
    }
    \put(68  , 1 ) { $\mathrm{m}_{\dipsi}$ }
    \put(118 , 1 ) { $\left[ \mathrm{GeV}/c^2 \right]$}
    \put(97,73) {  %
      $\begin{array}{l}                               %
	\text{LHCb}                 \\         %
        \sqrt{s} = 7~\mathrm{TeV}            %
      \end{array}$ } 
    \put(2,55) { 
      \begin{sideways}%
	$\frac{\mathrm{d}\sigma^{\dipsi}}{\mathrm{dm}_{\dipsi}} \left[ \frac{\mathrm{nb}}{1~\mathrm{GeV}/c^2}\right]$
      \end{sideways} 
    }
  \end{picture}
  \caption {
    Differential production cross-section for \jpsi~pairs
    as a function of the invariant mass of the \dipsi~system. The
    points correspond to the data. Only
    statistical uncertainties are included in the error
    bars. The~shaded area corresponds to prediction by the model 
    described in Ref.~\cite{Berezhnoy:2011xy}. 
  }
  \label{fig:dipsi_mass}
\end{figure}
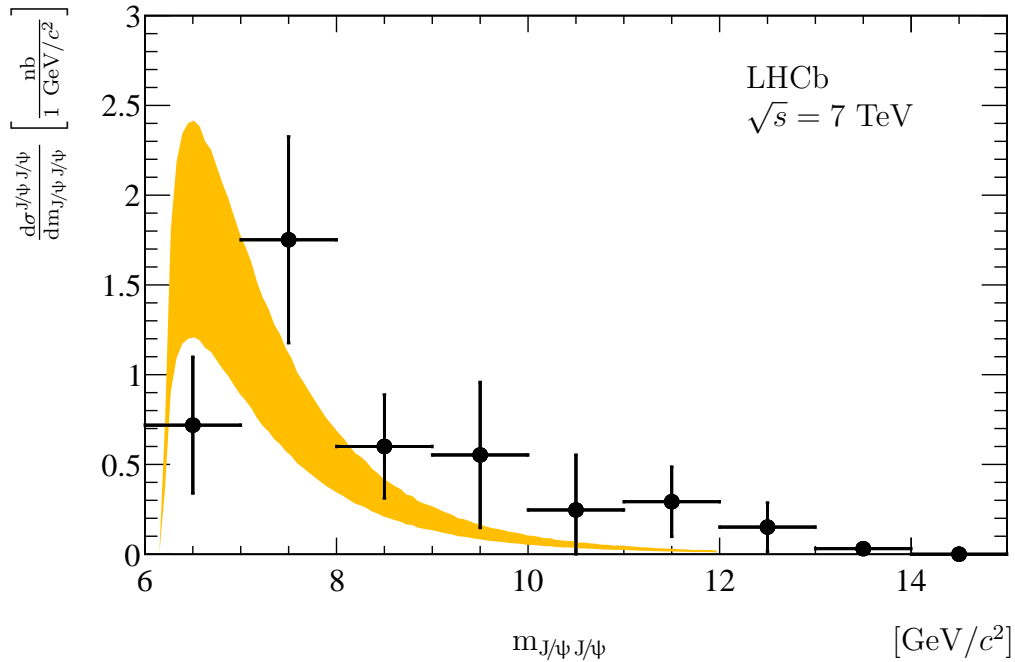

The differential production cross-section of \jpsi~pairs
as a function of the invariant mass of the \dipsi~system
is shown in Fig.~\ref{fig:dipsi_mass}.
The whole analysis chain has been repeated for each bin 
of \dipsi~invariant mass to get the differential production 
cross-section. The~bulk of the distribution 
is concentrated in the~low invariant mass region. 
A theoretical prediction for the shape of this distribution taking
into account both direct production and feeddown 
from $\Ppsi(2S)$~decays~\cite{Berezhnoy:2011xy} is overlaid. 
Within the available statistics the~agreement between the
data and the prediction is reasonable. 

%% file: summary.tex
%

\section{Conclusions}
The production of \jpsi~pairs 
in proton-proton collisions at a centre-of-mass 
energy of~$7~\mathrm{TeV}$  
has been observed with a statistical significance in excess of~$6\sigma$. 
The data are consistent with  the predictions given in 
Refs.~\cite{Qiao:2009kg,Berezhnoy:2011xy}.
The higher statistics 
that will be collected during the 2011 data-taking period will allow the 
kinematic properties of these events to be studied and different
production models to be probed.

%% file: acknowledgments.tex
\section*{Acknowledgments}

We would like to thank A.K.~Likhoded for many fruitful discussions.
We express our gratitude to our colleagues in the CERN accelerator
departments for the excellent performance of the LHC. We thank the
technical and administrative staff at CERN and at the LHCb institutes,
and acknowledge support from the National Agencies: CAPES, CNPq,
FAPERJ and FINEP (Brazil); CERN; NSFC (China); CNRS/IN2P3 (France);
BMBF, DFG, HGF and MPG (Germany); SFI (Ireland); INFN (Italy); FOM and
NWO (Netherlands); SCSR (Poland); ANCS (Romania); MinES of Russia and
Rosatom (Russia); MICINN, XuntaGal and GENCAT (Spain); SNSF and SER
(Switzerland); NAS Ukraine (Ukraine); STFC (United Kingdom); NSF
(USA). We also acknowledge the support received from the ERC under FP7
and the Region Auvergne.
%

%% file: references.tex

%% file: main.bbl
\begin{thebibliography}{99}
  
\bibitem{Kartvelishvili:1978id}
  V.G.~Kartvelishvili, A.K.~Likhoded and S.R.~Slabospitsky,
  ``D~meson and $\Ppsi$~meson production in hadronic interactions'',
  Sov.\ J.\ Nucl.\ Phys.\  {\bf 28}, 678 (1978)
  [Yad.\ Fiz.\  {\bf 28}, 1315 (1978)].

  
\bibitem{Berger:1980ni}
 E.L.~Berger and D.L.~Jones, 
 ``Inelastic photoproduction of \jpsi and $\PUpsilon$ by gluons'',
  Phys.\ Rev.\ {\bf{D23}}, 1521, (1981).

\bibitem{Baier:1981uk}
  R.~Baier and R.~Ruckl,
  ``Hadronic production of \jpsi and $\PUpsilon$: 
  transverse momentum distributions'',
  Phys.\ Lett.\  B {\bf 102}, 364 (1981).

  
\bibitem{Abe:1992ww}
  F.~Abe {\it et al.}, 
  ``Inclusive \jpsi, $\Ppsi(2S)$ and 
  $\mathrm{b}$-quark production in 
  $\bar{\mathrm{p}}\mathrm{p}$-collisions at 
  \mbox{$\sqrt{s}=1.8$}~TeV'',
  Phys.\ Rev.\ Lett.\  {\bf 69}, 3704 (1992). 
  
\bibitem{Braaten:1994vv}
  E.~Braaten and S.~Fleming,
  ``Color octet fragmentation and 
  the $\Ppsi^{\prime}$-surplus at the Tevatron'',
  Phys.\ Rev.\ Lett.\  {\bf 74}, 3327 (1995).
      
  

  
\bibitem{Campbell:2007ws}
  J.M.~Campbell, F.~Maltoni and F.~Tramontano,
  ``QCD corrections to \jpsi and $\PUpsilon$~production 
  at hadron colliders'',
  Phys.\ Rev.\ Lett.\  {\bf 98}, 252002 (2007).
  
\bibitem{Gong:2008sn}
  B.~Gong and J.X.~Wang,
  ``Next-to-leading-order QCD corrections to 
  \jpsi~polarization at Tevatron
  and Large-Hadron-Collider energies'',
  Phys.\ Rev.\ Lett.\  {\bf 100}, 232001 (2008).

\bibitem{artoisenet:2008}
  P.~Artoisenet, J.M.~Campbell, J.-P.~Lansberg, 
  F.~Maltoni and F.~Tramontano,
  ``$\PUpsilon$~production at Fermilab Tevatron and LHC energies'',
  Phys.\ Rev.\ Lett.\  {\bf 101}, 152001 (2008).

\bibitem{lansberg:2009}
  J.-P.~Lansberg, 
  ``On the mechanisms of heavy-quarkonium hadroproduction'',
   Eur.~Phy.~J. {\bf{C61}}, 693 (2009). 

\bibitem{Brambilla}
  N.~Brambilla {\it{et al.}}, 
  ``Heavy quarkonium: progress, puzzles, and opportunities'',
   Eur.~Phy.~J. {\bf{C71}}, 1534 (2011). 

\bibitem{Kom:2011bd}
  C.H.~Kom, A.~Kulesza and W.J.~Stirling,
  ``Pair production of \jpsi as a probe of double parton scattering at LHCb'',
   Phys.\ Rev.\ Lett.\ {\bf{107}}, 092002 (2011). 

\bibitem{Baranov:2011ch}
  S.P.~Baranov, A.M.~Snigirev and N.P.~Zotov,
  ``Double heavy meson production through double parton scattering in hadronic
  collisions'',
  \href{http://arxiv.org/abs/1105.6276}{arXiv:1105.6276 [hep-ph]}.
 
\bibitem{Novoselov:2011ff}
  A.~Novoselov,
  ``Double parton scattering as a source of quarkonia pairs in LHCb'',
  \href{http://arxiv.org/abs/1106.2184}{arXiv:1106.2184 [hep-ph]}.




\bibitem{Badier:1982ae}
  J.~Badier {\it et al.}, 
  ``Evidence for $\Ppsi\Ppsi$~production in $\pi^-$ 
  interactions at 150~GeV/$c$ and
  280~GeV/$c$'',
  Phys.\ Lett.\  B {\bf 114}, 457 (1982).
  
\bibitem{Badier:1985ri}
  J.~Badier {\it et al.}, 
  ``$\Ppsi\Ppsi$~production and limits 
  on beauty meson production from 400~GeV/$c$
  protons'',
  Phys.\ Lett.\  B {\bf 158}, 85 (1985).
    
\bibitem{Kartvelishvili:1984ur}
V.G.~Kartvelishvili and S.M.~Esakiya, 
``On hadron induced production of \jpsi meson pairs'',
Yad.\ Fiz.\ {\bf 38}, 722 (1983).

\bibitem{Humpert:1983yj}
  B.~Humpert and P.~Mery,
  `` $\Ppsi\Ppsi$~production at collider energies'',
  Z.\ Phys.\  C {\bf 20}, 83 (1983).
  
\bibitem{Kiselev:1988mc}
  V.V.~Kiselev, A.K.~Likhoded, S.R.~Slabospitsky and A.V.~Tkabladze,
  ``Hadronic pair production of $\Ppsi$ particles with large $M(\Ppsi\Ppsi)$,''
  Sov.\ J.\ Nucl.\ Phys.\  {\bf 49}, 1041 (1989) 
  [Yad.\ Fiz.\  {\bf 49}, 1681 (1989)].
  
\bibitem{Qiao:2009kg}
  C.F.~Qiao, L.P.~Sun and P.~Sun,
  ``Testing charmonium production mechanism via 
  polarized \jpsi-pair
  production at the LHC'',
  J.\ Phys.\ G {\bf 37}, 075019 (2010).
  
\bibitem{Berezhnoy:2011xy}
  A.V.~Berezhnoy, A.K.~Likhoded, A.V.~Luchinsky and A.A.~Novoselov,
  ``Double \jpsi~meson production at LHC and $4c$-tetraquark state'',
  arXiv:1101.5881 [hep-ph].
  
  

  
\bibitem{LHCb} 
  A.A.~Alves {\it et al.},
  ``The LHCb detector at LHC'',
  JINST {\bf 3}, S08005 (2008). 
  
\bibitem{pythia} 
T. Sj\"{o}strand, S. Mrenna and P. Z. Skands, 
  ``{\sc{Pythia}}~6.4 physics and manual'',
 JHEP {\bf{05}}, 026 (2006), version 6.422, 
\href{http://arxiv.org/abs/hep-ph/0603175}{arXiv:hep-ph/0603175}.

\bibitem{belyaev}
I. Belyaev {\it et al.},
 ``Handling of the generation of primary events in
   {\sc{Gauss}}, the LHCb simulation framework'',
  Nuclear Science Symposium Conference Record
  (NSS/MIC), IEEE, 1155  (2010).

\bibitem{evtgen} D.~Lange, 
  ``The {\sc{EvtGen}} particle decay simulation package'', 
  Nucl.~Instrum.~Meth.~A {\bf{462}}, 152 (2001).

\bibitem{geant} S. Agostinelli {\it et al.}, 
``{\sc{Geant4}}: a simulation toolkit'', 
 Nucl.~Instrum.~Meth.~A {\bf{506}}, 250 (2003).

\bibitem{Powell:2010zz}
 A.~Powell,
  ``Particle identification at LHCb'',
   PoS {\bf{ICHEP2010}}, 020 (2010).

\bibitem{Kullback} 
  S.~Kullback and R.A.~Leibler, 
``On information and sufficiency'', 
  Annals of Mathematical Statistics {\bf 22}, 79 (1951); \\
  S.~Kullback, 
  ``Letter to editor: the Kullback-Leibler distance'',
  The American Statistician {\bf 41}, 340 (1987).
  
  
\bibitem{Hulsbergen}
  W.D.~Hulsbergen,
  ``Decay chain fitting with a Kalman filter'',
  Nucl.~Instrum.~Meth.~A {\bf 552}, 566 (2005).

\bibitem{Gaiser} J.E.~Gaiser,
  ``Charmonium spectroscopy from radiative decays of the \jpsi and $\Ppsi^{\prime}$'',
  Ph.D.~Thesis, SLAC-R-255 (1982).

\bibitem{Skwarnicki} T.~Skwarnicki,
   ``A study of the radiative cascade transitions between the $\PUpsilon^{\prime}$~and 
    $\PUpsilon$~resonances'',
   Ph.D.~Thesis, DESY-F31-86-02 (1986).


\bibitem{Aaij:2011rj}
   R.~Aaij {\it{et al.}},
  ``Search for the rate decays 
$\mathrm{B}^0_{\mathrm{s}}\to\mumu$ and 
$\mathrm{B}^0_{\mathrm{d}}\to\mumu$'',
   Phys.\ Lett.\ B {\bf{699}}, 330 (2011).

\bibitem{Jpsi} 
   R.~Aaij {\it et al.},
   ``Measurement of \jpsi production in pp~collisions at
  $\sqrt{s}=7~\mathrm{TeV}$'', 
   Eur.~Phy.~J. C~{\bf{71}}, 1645 (2011). 
   
\bibitem{TisTos} 
  E.~L\'{o}pes~Azamar {\it et al.}, 
   ``Measurement of trigger efficiencies and biases'',
   \href{http://cdsweb.cern.ch/record/1152284}{CERN-LHCb-2008-073}, (2008).
 
\bibitem{K0S} 
  R.~Aaij {\it et al.},
  ``Prompt $\mathrm{K}^0_{\mathrm{S}}$ 
  production in $\mathrm{pp}$~collisions at 
  $\sqrt{s}=0.9~\mathrm{TeV}$'',
  Phys.~Lett.~B  {\bf 693}, 69 (2010).  


  

  

 
  
  
  
  
\bibitem{VanDerMeer}
  S.~van~der~Meer, ``Calibration of the effective beam height in the ISR'',
  ISR-PO/68-31 (1968).
  
\bibitem{BeamProfile}
  M.~Ferro-Luzzi,
  ``Proposal for an absolute luminosity determination in colliding beam experiments 
  using vertex detection of beam-gas interactions'', 
  Nucl.\ Instrum.\ Meth.~A {\bf 553}, 388 (2005).
  


\bibitem{Lumi} 
  G.~Anders {\it et al.,} 
  ``LHC bunch current normalization for the October 2010 luminosity calibration measurements'',
  CERN-ATS-Note-2011-016 PERF (2011).

\bibitem{Lumi2}
   R.~Aaij {\it et al.,}
  ``Absolute luminosity measurements with the LHCb detector at the LHC'',
   LHCB-PAPER-2011-015, CERN-PH-EP-2011-157, 
   \href{http://arxiv.org/abs/1110.2866}{arXiv:1110.2866},
   submitted to JINST.
  
\bibitem{PDG} 
  K.~Nakamura {\it et al.,} 
  ``Review of particle physics'',
  J.\ Phys.\ G {\bf 37}, 075021 (2010).
  

\end{thebibliography}
